# Hysteretic behavior of electrical conductivity in packings of particles

Behzad Ghanbarian[1*] and Muhammad Sahimi[2]

[1] Bureau of Economic Geology, Jackson School of Geosciences, University of Texas at Austin, Austin, Texas 78713, USA

[2] Mork Family Department of Chemical Engineering and Materials Science, University of Southern California, Los Angeles, California 90089-1211, USA

**Abstract** We address the problem of predicting saturation-dependent electrical conductivity $\sigma$ in packings of spheres during drainage and imbibition. The effective-medium approximation (EMA) and the universal power law of percolation for $\sigma$ are used, respectively, at higher and low water saturations to predict the conductivity, with the crossover between the two occurring at some intermediate saturation $S_{wx}$. The main input to the theory is a single parameter that we estimate using the capillary pressure data. The predictions are compared with experimental, as well as numerical data for three distinct types of packings. The results for drainage in all the packings indicate that the universal power law of percolation is valid o*ver the entire range of* $S_w$. For imbibition, however, the universal power law crosses over to the EMA at $S_{wx} = 0.5$. We also find that the effect of the pore-size distribution on the $\sigma$-$S_w$ relation is minimal during both drainage and imbibition.

* Corresponding author's email address: b.ghanbarian@gmail.com

**1 Introduction**

The relationship between the effective permeability *k* and electrical conductivity *σ* of porous media has been studied for decades (see, for example, Johnson et al. 1986; Katz and Thompson 1986,1987; Bernabe and Bruderer 1998; for a comprehensive review, see Sahimi 2011). Since *k* has the units of a length squared, its square root is a characteristic length scale of porous media, representing a static property. On the other hand, *σ* is a dynamic property that characterizes flow of electrons in a pore space saturated by a conducting fluid, such as brine. Thus, in principle, there is no *exact* relation between *k* and *σ*. Despite this, *σ* is commonly used in geophysics and hydrology to predict *k* and the relative permeabilities (see, for example, Revil and Cathles 1999; Doussan and Ruy 2009; Niu et al. 2015; Mawer et al. 2015). For this reason accurate prediction of *σ* has been a problem of great interest for a long time (see e.g., Knight 1991; Friedma, 2005; Zhan et al. 2010; Revil 2016), particularly under *partially-saturated* conditions, for which various models have been proposed to predict *σ*. Among them is the empirical Archie's law (Archie 1942), which is valid only for water-wet rocks deprived of microporosity or substantial clay-exchange cations (Toumelin and Torres-Verdín 2008), and has been extensively used to determine *σ*. A recent review of the theoretical models of *σ* is given by Ghanbarian et al. (2015a).

Mualem and Friedman (1991) represented the pore space by a bundle of series-parallel capillary tubes (Mualem 1976) in order to predict *σ*. By including the concept of residual volumetric water content, they proposed power laws for *σ* under fully- and partially-saturated conditions. The exponents that characterized their power laws were 1.5 and 2.5 [see their Eqs. (30) and (31)] for, respectively, the fully- and partially-saturated media. The effect of hysteresis, pore shape, pore-size distribution (PSD), viscosity, and wettability on saturation-dependent $\sigma$ has been well studied using pore-network models (PNMs) and/or solving the Laplace equation in digitized images (see Schwartz and Kimminau 1987; Tsakiroglou and Fleury 1999a,b; Man and Jing, 2000; Bekri et al., 2003). Bryant and Pallatt (1996) argued that $\sigma$ is dependent upon the geometry of the pore space, and proposed a PNM for computing it for a packing of spheres. The predictions were in good agreement with experimental data. Suman and Knight (1997) studied the effect of the pore structure and wettability on saturation-dependence of $\sigma$ during drainage and imbition by means of a PNM. The effect of the wettability was investigated, and the influence of the pore structure was studied by varying the breadth of the PSD and the pore-size correlations. Suman and Knight (1997) reported that the effect of hysteresis is more significant in oil-wet porous media than in the water-wet ones. Hysteresis in both types of porous media decreased, however, as the extent of the correlation between the pores increased. Li et al. (2015) computed $\sigma$ during drainage and imbition, modeled by invasion percolation (Chandler et al. 1982; Wilkinson and Willemsen 1983, Knackstedt et al. 2000) in the PNMs under completely water-wet conditions. The computed $\sigma$ did not generally follow Archie's law (consistent with the experimental data of Longeron et al. 1989), and exhibited either downward or upward curvature at lower saturations, depending on whether the effect of the thin water films was included in the model.

Percolation theory has been also invoked to predict $\sigma$ in partially-saturated soil and rock (see, for example, Heiba et al. 1992; Zhou et al. 1997; Wang et al. 2007; Montaron 2009). Using percolation theory, Ewing and Hunt (2006) proposed the following equation for the saturation-dependence of $\sigma$,

$$\sigma_r = \frac{\sigma(S_w)-\sigma_s}{\sigma(S_w=1)-\sigma_s} = \left[\frac{S_w-S_{wc}}{1-S_{wc}}\right]^t, \quad S_{wc} \leq S_w \leq 1 \tag{1}$$

where $S_w$ and $S_{wc}$ are, respectively, the water saturation and its critical value (or the percolation threshold), $\sigma(S_w)$ and $\sigma(S_w=1)$ are the electrical conductivities under partially- and fully-saturated conditions, $t = 2$, and $\sigma_s$ is the surface conductivity. Equation (1), which is similar in form to the model of Mualem and Friedman (1991), assumes that the solid and bulk phases conduct strictly in parallel. The exponent $t = 2$ in Eq. (1) incorporates the effect of pore connectivity, tortuosity and the correlations that arise as a result of tortuous conduction paths near the critical saturation.

The power laws of Mualem and Friedman (1991) and Ewing and Hunt (2006) are not, however, valid over the entire range of saturation, particularly if the porous medium is highly heterogeneous. To address such cases, Ewing and Hunt (2006) proposed another power law in which, instead of the universal $t = 2$ in Eq. (1), a non-universal exponent was used. The model invoked the critical-path analysis (CPA), which is based on percolation theory and included a crossover saturation $S_{wx}$ at which the universal power law, Eq. (1), switches to a non-universal one based on the CPA, given by

$$\sigma_r = \frac{\sigma(S_w)-\sigma_s}{\sigma(S_w=1)-\sigma_s} = \left[\frac{1/\phi-1+S_{wx}-S_{wc}}{1/\phi-S_{wc}}\right]^{\frac{1}{3-D}} \left[\frac{S_w-S_{wc}}{S_{wx}-S_{wc}}\right]^2, \quad S_{wc} \leq S_w \leq S_{wx} \tag{2a}$$

$$\sigma_r = \frac{\sigma(S_w)-\sigma_s}{\sigma(S_w=1)-\sigma_s} = \left[\frac{1/\phi-1+S_w-S_{wc}}{1/\phi-S_{wc}}\right]^{\frac{1}{3-D}}, \quad S_{wx} \leq S_w \leq 1 \tag{2b}$$

where $\phi$ is the total porosity and $D$ is the pore space fractal dimension. Equations (2) are valid for porous media whose PSD is broad or, equivalently, when $D \geq 3 - \frac{(1-S_{wc})}{2(1/\phi-S_{wc})}$. For example, in a porous medium with $\phi = 0.3$ and $S_{wc} = 0.25$, Eq. (2) is applicable only if $D \geq 2.88$. By comparing the predictions of Eqs. (1) and (2) with the experimental data for sediments and soils, Ewing and Hunt (2006) argued that Eq. (1) should predict accurately the saturation-dependence of $\sigma$.

Ghanbarian et al. (2014; see, however, Dashtian et al, 2015 for a note of caution) combined the universal power law of percolation, i.e., $\sigma_r \sim (S_w - S_{wc})^t$ in which $t = 2$, with the effective-medium approximation (EMA), $\sigma_r \sim (S_w - 2/Z)$, in which $Z$ is the average pore connectivity, to obtain,

$$\sigma_r = \frac{\sigma(S_w)}{\sigma(S_w=1)} = \frac{S_{wx}-2/Z}{1-2/Z}\left[\frac{S_w-S_{wc}}{S_{wx}-S_{wc}}\right]^2, \quad S_{wc} \leq S_w \leq S_{wx} \tag{3a}$$

$$\sigma_r = \frac{\sigma(S_w)}{\sigma(S_w=1)} = \frac{S_{wx}-2/Z}{1-2/Z}, \quad S_{wx} \leq S_w \leq 1 \tag{3b}$$

The significance of a crossover between the EMA and percolation description of the conductivity was first pointed out by Sahimi et al. (1983b). By comparing the predictions with the data for sandstones and carbonates, Ghanbarian et al. (2014) found that, under fully-saturated conditions, the crossover point depends on the pore space morphology. Equation (3) was also successfully used to predict saturation-dependent gas and solute diffusivities (Hunt et al. 2014a,b; Ghanbarian and Hunt 2014; Ghanbarian et al. 2015b) in rocks, soils, and packings of spheres, and is consistent with random walk results of saturation-dependent diffusion simulation in isotropic and homogeneous reconstructed unimodal porous media reported by Valfouskaya and Adler (2005). Whether it can also predict $\sigma$ in partially-saturated packings of mono-sized spheres and glass beads during drainage and imbibition is still an open question.

The main objectives of this study are, therefore, (1) investigating hysteresis in the $\sigma$ -$S_w$ relation in packings of mono-sized spheres; (2) evaluating the accuracy of Eq. (3) against numerical and experimental data for such packings, and (3) studying the effect of the PSD on the saturation-dependence of $\sigma$. Note that we study the saturation dependence of static, low-frequency electrical conductivity in which such processes as polarization and relaxation are negligible. The general frequency-dependent dynamic conductivity may be studied by the EMA of Sahimi et al. (1983a).

## 2 Theory

The predictions that we present in this paper rely on two distinct theories that we describe first.

### 2.1 The pore-size and pore-conductance distributions

The PSD of the pore space of a packing of spheres is well approximated by a power law (Bryant et al. 1993). Indeed, rigorous analysis by Halperin et al. (1985) and Feng et al. (1987) indicated that the pore-conductance distribution (PCD) of the same packing is of the power-law type. On the other hand, the electrical conductance $g$ of a cylindrical pore of radius $r$ and length $l$, filled with a fluid of conductivity $\sigma_f$ conforms to

$$g = \sigma_f \frac{\pi r^2}{l} \tag{4}$$

Clearly, then, if the PCD follows a power law, so also does the PSD. Since the pore space of packings of spheres is isotropic, the conductance of the pores scales with a single scaling factor in all the directions, and we assume that, $l \propto r$ (although others have argued that $l \propto 1/r$). Accordingly, Eq. (4) becomes

$$g \propto r^\gamma \tag{5}$$

in which $\gamma = 1$, if $l \propto r$. In the classical percolation problem, the various variables, such as the pores' conductance, take on discrete values. But, if they are distributed according to a continuous probability distribution function, then the problem is referred to as *continuum percolation*. The analyses of Halperin et al. (1985) and Feng et al. (1987) demonstrated that the conductance distribution $f(g)$ of the pore space of a packing of spheres is given by

$$f(g) = c_g g^{-\alpha}, \quad g_{min} \leq g \leq g_{max} \tag{6}$$

in which $c_g = (1-\alpha)/(g_{max}^{1-\alpha} - g_{min}^{1-\alpha})$ and $\alpha = (\gamma - 1)/\gamma$.

Using the EMA, Kogut and Straley (1979) studied steady-state conduction, and Sahimi et al. (1983a) investigated frequency-dependent diffusion and conduction in disordered materials and media. Both groups showed that power law (1) that describes $\sigma$ near the percolation threshold is characterized by a *non-universal* exponent $t$, if $0 < \alpha < 1$ (i.e. one that depends on the details of the PCD), whereas its universality is restored for $\alpha \leq 0$. Using a scaling analysis, Halperin et al. (1985) and Feng et al. (1987) made the approximate work of Kogut and Straley (1979) and Sahimi et al. (1983a) rigorous. Therefore, in porous media in which the PCD is given by a power law, one should expect $\sigma$ to follow Eq. (1) near the percolation threshold with a universal $t = 2$, if $\gamma \leq 1$, and a non-universal $t$ for $\gamma > 1$. In the latter case (Straley 1982; Feng et al. 1987),

$$t = max\left\{2, 1 + \nu + \frac{\alpha}{1-\alpha}\right\} = max\{2, 1 + \nu + (\gamma - 1)\} \tag{7}$$

where $\nu$ is the exponent that characterizes the divergence of the percolation correlation length near the percolation threshold, with $\nu \approx 0.88$ (Stauffer and Aharony 1994). Thus, so long as $\gamma \geq 1.12$, one has

$$t = 0.88 + \gamma \tag{8}$$

The significance of the non-universal power laws of percolation to correct comparison of theoretical predictions for flow and transport in porous media with the experimental data was recently discussed and emphasized by Sahimi (2012).

**2.2 Pore-size distribution and the capillary pressure**

Next, we need a relation between the PSD and the capillary pressure-saturation curve. Many porous media have a fractal pore space (Katz and Thompson 1985; Perrier et al. 1996) with their PSD given by,

$$f(r) = \frac{D}{r_{min}^{-D} - r_{max}^{-D}} r^{-1-D}, \quad r_{min} \leq r \leq r_{max} \tag{9}$$

where $r_{min}$ and $r_{max}$ are the minimum and maximum pore radii, and $D$ is the fractal dimension of the pore space. If $-1 < D < 3$, one has a pore space in which smaller pores are more probable than the larger ones. If $D = -1$, Eq. (9) reduces to a uniform PSD. If $D < -1$, however, larger pores are more probable than the smaller ones. Using the relationship among porosity, $r_{min}/r_{max}$, and $D$, Ghanbarian-Alavijeh and Hunt (2012) demonstrated that negative values of $D$ is permissible theoretically [see their Eq. (7) and Table 1; see also Mandelbrot 1990], although such values are associated with very complex structures.

The pore space of a packing of spheres does not have a fractal structure, unless the porosity is so low that the pore space is at or very close to the percolation threshold or the critical porosity. In most porous media, $D > 2$ and is a measure of ruggedness of the pore space and its pore surface. Thus, since the pore space of a packing of sphere has a power-law PSD similar to Eq. (9), we refer to such porous media as *power-law porous media*, rather than fractal porous media. If $D < 2$, we view $D$ merely as a parameter of the PSD (or PCD, as the PSD and PCD are related). As we demonstrate below, for all the packings that we study, $D < 2$, justifying our view of $D$.

Following Eq. (9), the capillary pressure curve for a power-law pore space can be derived. By invoking the Young-Laplace equation for the capillary pressure, $P_c = A/r$ in which $A$ is a constant, one has

$$S_w = 1 - \frac{\beta}{\phi}\left[1 - \left(\frac{P_c}{P_e}\right)^{D-3}\right], \quad P_e \leq P_c \leq P_{cmax} \tag{10}$$

where $P_e$ is the entry (displacement) pressure at which the invading fluid (e.g., air) begins displacing the defending fluid (water, for example), $P_{cmax} = A/r_{min}$ is the maximum pressure, and $\beta = \phi r_{max}^{3-D}/(r_{max}^{3-D} - r_{min}^{3-D})$ and is smaller for a pore space with more uniform pore sizes. Equation (10) reduces to the models proposed by Tyler and Wheatcraft (1990) and Rieu and Sposito (1991) for $\beta = \phi$ and 1, respectively. It is also equivalent to the empirical Brooks and Corey (1964) model when $\beta = \phi$ and $\lambda = 3 - D$ (Perrier et al. 1996) in which $\lambda$, referred to by Brooks and Corey as the PSD index, may take on any positive value, being small for media with a wide range of pore sizes and large for a pore space with a relatively uniform PSD. Note that $b = f$ requires that $r_{min} \to 0$, implying a very broad PSD ($r_{max}/r_{min} \to \infty$). Recall that such broad PSDs are the cause of the non-universality of the conductivity exponent $t$.

Thus, if we view Eq. (10) as the saturation-capillary pressure relation for porous media in which the PSD and PCD are of power-law type, we can utilize it to study electrical conductivity of partially-saturated packings of spherical particles.

**3 The Data**

**3.1 Data of Mawer et al.**

We first utilized the database reported by Mawer et al. (2015) who carried out numerical computation of saturation-dependent capillary pressure and $\sigma$ for the Finney packing (Finney, 1970) whose porosity $\phi$ is 0.362, and 14 other packings with porosities $0.23 < \phi < 0.46$. Finney constructed random packings of spherical particles experimentally and recorded the coordinates of their centers. Salient details of each sphere packing are presented in Table 1. It has been used for simulating various flow and transport (Roberts and Schwartz 1985; Bryant et al. 1993) in porous media, and biological reaction and transport (Dadvar and Sahimi 2003) in packed-bed reactors, as its structure is very well characterized.

In order to generate a packing with a partial saturation, Mawer et al. (2015) saturated the pore space and carried out drainage and imbibition simulations to capture hysteresis between the two. The conductivity of the packings was computed using the finite-element approach (Garboczi 1998). Complete details are given by Mawer et al. (2015). To predict the saturation-dependence of $\sigma$, the parameters of the power-law model of the capillary pressure curve, namely, $D$, $b$, and $P_e$, were estimated by fitting Eq. (10) to the capillary pressure data computed by computer simulations.

**3.2 Data of Knackstedt et al.**

Knackstedt et al. (2007) studied conduction in a monodisperse glass bead pack with a porosity of 0.258, measured via porosimetry. The sample's image was obtained by micro-computed tomography with a resolution of 55 microns per voxel and a total of $2048^3$ voxels. They use a finite-element method to solve the Laplace equation under non-periodic boundary conditions by minimizing the energy using the conjugate-gradient method (Vaez Allaei and Sahimi, 2005), and calculated the electrical conductivity of the glass bead during drainage. Knackstedt et al. (2007) considered two limiting cases of wettability: strongly water-wet and strongly oil-wet. As expected, the non-wetting fluid resided in the regions with large covering spheres (i.e. large pores between the glass beads), while the wetting fluid was distributed in the small radius region.

**3.3 Data of Sharma et al.**

Sharma et al. (1991) used packings of glass beads the diameters of which varied from 0.18 to 0.25 mm. The beads were packed inside a resistivity cell equipped with two end-current electrodes and two ring-voltage electrodes, across which the electrical resistivity was measured using a four-electrode method. After packing the beads, an axial stress of 100 psi was applied, followed by a radial stress of 300 psi. The packings were then saturated using a 30,000 ppm brine solution. Sharma et al. (1991) used two non-wetting phase fluids, namely, n-decane, and naphtha, and measured the electrical resistivity at various water saturations during both drainage and imbibition. The digitized data used in the present study were taken from their Figures 6 and 8.

**4 Results and Discussions**

**4.1 Predictions for the Data of Mawer et al.**

We first discuss the capillary pressure curve, and the pore space parameters $D$ and $b$ for the packings under drainage and imbibition conditions, after which we compare the predictions for $\sigma$ with the data for the 15 packings.

**4.1.1 Capillary pressure: drainage vs. imbibition**

The estimated parameters $D$, $b$, and $P_e$ are listed in Table 1, indicating that $-1.64 < D < 0.62$ for drainage and $0.99 < D < 1.74$ for imbibition. The latter range is compatible with $D = 1.34$ that we derived based on the PSD of the Finney packing, reported by the Dadvar and Sahimi (2003). A comparison of $D$ for drainage and imbition reported in Table 1 indicates that the imbition PSD is broader than that of drainage, in agreement with the general results of Ojeda et al. (2006) and consistent with the assumption of $D = 0.5$ ($\lambda = 2.5$) and 1.75 ($\lambda = 1.25$), respectively, for drainage and imbibition (Gerhard et al. 2007; see their Table 1). We should point out that the negative values of $D$ presented in Table 1 are in accord with the large values of the PSD index $\lambda$ of the Brooks and Corey (1964) for packings of spheres. Brooks and Corey (1964), Mualem (1976), Eckberg and Sunada (1984), and Ioannidis et al. (2006) reported, respectively, that $\lambda = 7.3$, 6.24, 7, and 5.07 corresponding to $D = -4.3$, -3.24, -4, and -2.07 [if $\beta = \phi$ in Eq. (5)] for glass beads of uniform sizes.

**4.1.2 Electrical conductivity during drainage**

Figure 1 presents the computed capillary pressure and $\sigma_r$ during drainage and their dependence on the water saturation. Although the data are scattered, exhibiting relatively wide range of the PSDs, $\sigma_r$ follows the same trends with the water saturation, indicating clearly that the effect of the PSD and PCD on $\sigma_r$ is minimal. Since the packings are relatively homogeneous, their pore space is well connected, and the spheres' surface is smooth, the water saturation vanishes at a non-zero capillary pressure. That is, the irreducible water saturation is zero; see Figure 1(a). We, therefore, set $S_{wc} = 0$ to predict the saturation-dependence of $\sigma_r$ using Eq. (3).

For all the packings, Eq. (3) reduces to the universal power law of percolation, implying that $S_{wx} = 1$. We find that Eq. (1) with $\sigma_r = [(S_w - S_{wc})/(1 - S_{wc})]^2$ and $S_{wc} = 0$ is accurate over the *entire range of water saturation*; see Figure 1(b). This is consistent with Feng et al. (1987) who pointed out that when the exponent $\alpha < 0$ in Eq. (6), the effect of the PCD on $\sigma$ is negligible. Ewing and Hunt (2006) and Ghanbarian et al. (2014; 2015a) emphasized also that one should expect the saturation-dependence of $\sigma$ to follow universality, if the medium is not too heterogeneous (i.e. if $D < 2$).

Kogut and Straley (1979), Sahimi et al. (1983a), and Feng et al. (1987) all emphasized that the propensity of very small or zero conductance is the root cause of the non-universality of the exponent $t$. Putting it another way, $r_{min} = 0$ is the origin of non-universality (Sahimi et al. 1983a). Although we find $b = f$ for several of the packings (see Table 1), their $\sigma_r$ conforms to the universal power law, Eq. (1) with $t = 2$; see Figure 2(b). That for all the packings $D < 2$ (Table 1) implies that in the present formulation the key parameter for the universality or non-universality of $t$ is $D$: $\sigma_r$ follows the universal power law of percolation, if $D < 2$. Ghanbarian et al. (2015a) discussed the special case $b = f$ using the CPA, and pointed out that there exists no criterion to distinguish the crossover point separating the power law from the CPA when $\beta = \phi$. They proposed combining the two power laws to model $\sigma_r$ over the entire range of saturation, if $D \geq 2$.

**4.1.3 Electrical conductivity during imbibition**

Figure 2 presents the computed capillary pressure and $\sigma_r$ for all the packings during imbibition. Although, similar to drainage, the data follow a wide range of trends, $\sigma_r$ conforms to the same behavior as in drainage: varying linearly with water saturations near 1, but following a quadratic dependence on the saturation when it is close to $S_{wc}$. The linear behavior is consistent with the EMA, Eq. 3(b), while the quadratic dependence is compatible with the universal power law of percolation, Eq. 3(a). The computed results are in accord with the universal electrical conductivity model of Ghanbarian et al. (2014), Eq. (3). We find that, with $S_{wc} = 0$, $S_{wx} = 0.5$, and $Z = 8$ Eq. (3) provides predictions that are in very good agreement with the imbibition $\sigma_r$; see Figure 2(b). The crossover saturation that we identify, $S_{wx} = 0.5$, should be compared with 0.75, reported by Kiefer et al. (2009), while $Z = 8$ is consistent with the range of connectivities reported by Makse et al. (2000), Aste et al. (2006), Bhattad et al. (2011) and Jiang et al. (2016) for packings of spheres. As was the case for the drainage, the effect of the PSD and PCD on $\sigma_r$ is negligible.

Ghanbarian et al. (2015b) also showed that Eq. (3) was in agreement with gas and solute diffusion in mono-size packings of overlapping or non-overlapping spheres, in accord with Einstein's relation that $\sigma$ is proportional to the diffusion coefficient. They demonstrated that, as the PSD becomes very narrow, one should expect the crossover point between the EMA and the power-law regimes to occur below the numerical value of the corresponding porosity. This is consistent with the imbibition results. Bearing this in mind, one may expect the presence of a crossover point on the drainage $\sigma_r$ curve, since the drainage PSD is narrower than the imbibition, as quantified by the $D$ values in Table 1.

Figure 3(a) shows the imbibition and drainage capillary pressures for packing number 3 from Mawer et al. (2015). The maximum drainage capillary pressure, $P_{cmax} = 7.6$ m, is four times greater than $P_{cmax} = 1.9$ m for imbibition. Therefore, it may appear that the drainage PSD might be broader than the imbibition's. In contrast, for imbibition, $D = 1.74$ is greater than that of drainage, $D = 0.49$, by a factor of 3.5. In fact, in order to graphically compare the breadth of the PSD during drainage and imbibition, one should normalize the capillary pressures by the entry pressure $P_e$. Given that the entry pressure $P_e = 1.87$ and 0.29 m, we obtain, $P_{cmax}/P_e = r_{max}/r_{min} = 4.05$ and 6.54 for drainage and imbibition, respectively. The normalized capillary pressure curves in Figure 3(b) also confirm that the imbibition PSD is broader than that of the drainage. One should bear in mind, however, that the drainage PSD in other types of porous media may be broader than that of imbibition.

The dependence of $\sigma_r$ on water saturation and capillary pressure during drainage and imbibition for packing number 3 are shown in Figures 3(c) and 3(d). As expected, the broader the PSD, the sharper is the $\sigma_r$ curve. Therefore, one may expect the saturation-dependent $\sigma_r$ during imbibition to lie below the drainage, since the imbibition PSD is broader than that in drainage. As Figure 3(c) indicates, however, the imbibition $\sigma_r$ curve falls above the drainage, consistent with experimental measurements of Bourget et al. (1958), Longeron et al. (1989), and Knight (1991). We should, however, keep in mind that the water saturation level under drainage is not comparable with that under imbibition.

We show in Figure 3(d) the variations of $\sigma_r$ with the capillary pressure for during drainage and imbibition. At a specific capillary pressure, e.g. $P_c = 1$ m, where the drainage water saturation is greater than that in imbibition $S_w$ [see Figure 3(a)], the imbibition $\sigma_r$ is smaller than the drainage, which is expected since small pores are filled during imbibition, whereas large pores are occupied during drainage.

We point out that Knight (1991) investigated experimentally the hysteresis in the saturation-dependence of $\sigma_r$ in three sandstone samples, reporting pronounced hysteresis in the electrical resistivity (inverse of the conductivity) throughout much of the saturation range, particularly at intermediate water saturations. This agrees with Figure 3(c). She showed that in the hysteric (intermediate saturation) region, the electrical resistivity measured during imbibition was consistently less than that measured during drainage. The resistivity in most of the sandstones studied by Knight (1991) tended, however, to a finite value at zero water saturation ($S_w = 1$), indicating that surface conduction contributed to flow at completely dry conditions. We also point out that the data reported by Knight (1991) had been measured at 60 kHz, which is above the frequency range affected by polarization at the sample/electrode interface, as well as far above the low-frequency range addressed in this study. Nonetheless, our predictions are in reasonable agreement with the data of Knight (1991) for sandstones.

**4.2 Predictions for the Data of Knackstedt et al.**

Figure 4 compares the electrical conductivity of the mono-sized glass bead during drainage, computed by Knackstedt et al. (2007) using the image of the packings, with the predictions of Eq. (1) under strongly water- and oil-wet conditions. Following the results that we described for the data of Mawer et al. (2015), we set $S_{wc} = 0$, $\sigma_s = 0$, and $t = 2$ and found very good agreement between the water-wet numerical simulations of Knackstedt et al. (2007) and the predictions of Eq. (1). This confirms that the effect of the PSD in water-wet sphere packings is negligible, and that the universal power law of percolation, Eq. (1), predicts accurately the saturation-dependence of the electrical conductivity during drainage. Equation (1) with $S_{wc} = 0$, $\sigma_s = 0$, and $t = 2$ overestimated the saturation-dependence of the electrical conductivity under oil-wet conditions. Given that the morphology of the pore space of the glass bead packings of Knackstedt et al. (2007) is the same for both water- and oil-wet conditions, Fig. 4 clearly indicates the effect of the wettability on the saturation-dependence of the electrical conductivity during drainage. Although we found better agreement between the predictions of Eq. (1) with $S_{wc} = 0.3$, $\sigma_s$

= 0, and $t = 2$ (not shown) and Knackstedt et al.'s computed conductivity of the oil-wet packings, the sharp change in the $\sigma_r - S_w$ functional form presented in Fig. 4(b) may not be attributed to a large critical water saturation, because Morrow (1970) demonstrated experimentally that the irreducible wetting-phase saturation varied slightly with the change in the contact angle in random packings of 3-mm teflon spheres. More specifically, Morrow (1970) reported irreducible wetting-phase saturations in the range 0.066 - 0.09 with no specific trend of variations as the contact angle was varied from 0 to 108°. Since within the framework of percolation theory the value of conductivity exponent $t$ depends only on the pore-size or the conductance distribution [see Eq. (7)], the conclusion is that one should expect the proposed theoretical method in this study to predict accurately the saturation-dependence of the electrical conductivity in water-wet porous media. Further investigation is required to address the possible effect of the wettability on the exponent $t$ in Eq. (1).

**4.3 Predictions for the Data of Sharma et al.**

We show in Figure 5 the experimental data of Sharma et al. (1991) for the electrical conductivity during both drainage and imbibition. There is no remarkable difference between the two cases, and the effect of hysteresis on the saturation-dependence of the electrical conductivity in uniform packing of water-wet glass beads is negligible. Figure 5 also shows the predictions of Eq. (1), indicating good agreement with the experimental data. Due to lack of extensive experimental data, we only compared the predictions of Eq. (1) with such measurements, since precise determination of the crossover saturation in Eq. (3) requires accurate data (experimental or simulated) for the saturation-dependence of the electrical conductivity.

By fitting Archie's law, i.e., $\frac{\sigma_b}{\sigma_f} = S_w^n$, to their data, Sharma et al. (1991) reported that the average values of *n* are, respectively, 2.22 and 1.69 for the water-wet uniform glass beads during drainage and imbibition. Note that $n = 2.2$ reported for drainage is only 10 percent greater than the universal exponent of 2 predicted by percolation theory for the electrical conductivity, Eq. (1). The estimate is also consistent with that obtained by the numerical simulations of Mawer et al. (2015) shown in Fig. 3(c).

On the other hand, the exponent 1.69 reported by Sharma et al. (1991) for imbibition represents an intermediate value between Eq. (3) in which the electrical conductivity scales with the water saturation with the exponent 2 (as predicted by percolation theory) at saturations near the critical, and 1 that the EMA predicts near the fully-saturated condition. In other words, because the two regimes were mixed by Sharma et al. (1991), the resulting exponent represents an *effective, but not true, estimate* of the exponent. Sharma et al. (1991) stated that, "Two-endpoint *n* measurements indicate that for both glass beads and Berea cores the imbibition *n* values are smaller than those for drainage (Tables 1, 3). For water-wet beads, the two-point drainage-saturation exponents were higher than for the imbibition cycle by 27% (2.22 versus 1.69). Likewise, for water-wet cores, the average drainage *n* value was substantially higher than that for imbibition (1.53). These observations are consistent and reproducible and are also consistent with the observations of Swanson (1980), who reported similar differences." But, in fact, the emphasis of the present work is that, at least for imbibition, the two regimes should not be mixed.

Although we addressed the saturation-dependence of the electrical conductivity in packings of spheres and glass beads by means of percolation theory and the EMA, further investigation is required to study the possible effect of other factors and the circumstances under which the crossover saturation $S_{wx}$ may be altered.

## 5 Summary

This paper combined the EMA with the power law of percolation to predict the saturation-dependent electrical conductivity $\sigma_r$ in packings of particles. The predictions were evaluated against the numerical data for 15 packings reported by Mawer et al. (2015), as well as two other datasets by Knackstedt et al. (2007) and Sharma et al. (1991). By analyzing the capillary pressure data reported by Mawer et al. (2015), we showed that the imbibition PSD is broader than that of drainage. The estimated power-law exponent of the PSD, $D < 2$, indicated that the electrical conductivity should follow the universality of the conductivity exponent. Although the capillary pressure data varied remarkably among the 15 packings, their saturation-dependent electrical conductivity followed the same trends during both drainage and imbibition. This means that in the packings that we considered in which all the spheres have the same size, the effect of the PSD on the $\sigma_r$-$S_w$ relation is minimal. Our results also indicated that the drainage $\sigma_r$ conforms to the universal power law of percolation conductivity *over the entire range of water saturation*. The imbibition $\sigma_r$, however, exhibited a crossover from the universal power law to the EMA. Our predictions for the dataset of Knackstedt et al. (2007), computed by the universal power law of percolation theory, were accurate only for water-wet packing of mono-sized glass beads. We found the same trends for the Sharma et al. (1991) dataset.

**Acknowledgements**

B. G. is grateful to Chloe Mawer, Silicon Valley Data Science, for providing the results of numerical simulation of drainage and imbibition used in this study, and to Hugh Daigle, the University of Texas at Austin, for his comments on the very first draft of this paper. Publication was authorized by the Director, Bureau of Economic Geology.

Table 1. The capillary pressure curve fractal model parameters for 15 mono-sized sphere packs from Mawer et al. (2015).

| Pack | Porosity | Condition | $\beta$ | $D$ | $P_e$ (m) | $R^2$ |
|---|---|---|---|---|---|---|
| Finney | 0.36 | Drainage | 0.36 | -0.16 | 0.04 | 0.998 |
| | | Imbibition | 0.37 | 1.12 | 0.03 | 0.998 |
| 1 | 0.33 | Drainage | 0.36 | -0.09 | 1.38 | 0.998 |
| | | Imbibition | 0.37 | 1.68 | 0.22 | 0.993 |
| 2 | 0.35 | Drainage | 0.351 | -0.69 | 1.37 | 0.997 |
| | | Imbibition | 0.37 | 1.44 | 0.09 | 0.996 |
| 3 | 0.23 | Drainage | 0.24 | 0.49 | 1.87 | 0.999 |
| | | Imbibition | 0.26 | 1.74 | 0.29 | 0.997 |
| 4 | 0.25 | Drainage | 0.26 | 0.62 | 1.86 | 0.996 |
| | | Imbibition | 0.28 | 1.67 | 0.29 | 0.996 |
| 5 | 0.26 | Drainage | 0.26 | -0.58 | 1.84 | 0.999 |
| | | Imbibition | 0.28 | 1.60 | 0.29 | 0.997 |
| 6 | 0.28 | Drainage | 0.28 | -0.82 | 1.80 | 0.998 |
| | | Imbibition | 0.30 | 1.49 | 0.28 | 0.998 |
| 7 | 0.30 | Drainage | 0.30 | -1.41 | 1.83 | 0.997 |
| | | Imbibition | 0.32 | 1.38 | 0.28 | 0.998 |
| 8 | 0.32 | Drainage | 0.32 | -1.29 | 1.75 | 0.996 |
| | | Imbibition | 0.35 | 1.65 | 0.25 | 0.993 |
| 9 | 0.34 | Drainage | 0.34 | 0.21 | 1.51 | 0.992 |
| | | Imbibition | 0.36 | 1.53 | 0.25 | 0.993 |
| 10 | 0.36 | Drainage | 0.36 | -0.53 | 1.49 | 0.999 |
| | | Imbibition | 0.39 | 1.65 | 0.23 | 0.991 |
| 11 | 0.38 | Drainage | 0.38 | -0.63 | 1.45 | 0.998 |
| | | Imbibition | 0.39 | 1.27 | 0.24 | 0.995 |
| 12 | 0.42 | Drainage | 0.42 | -0.36 | 1.25 | 0.993 |
| | | Imbibition | 0.44 | 1.30 | 0.22 | 0.995 |
| 13 | 0.44 | Drainage | 0.44 | -0.61 | 1.25 | 0.999 |
| | | Imbibition | 0.46 | 1.17 | 0.22 | 0.995 |
| 14 | 0.46 | Drainage | 0.46 | -1.64 | 1.24 | 0.996 |
| | | Imbibition | 0.47 | 0.99 | 0.21 | 0.995 |
| Average | 0.34 | Drainage | 0.34 | -0.50 | 1.46 | 0.997 |
| | | Imbibition | 0.36 | 1.45 | 0.23 | 0.995 |

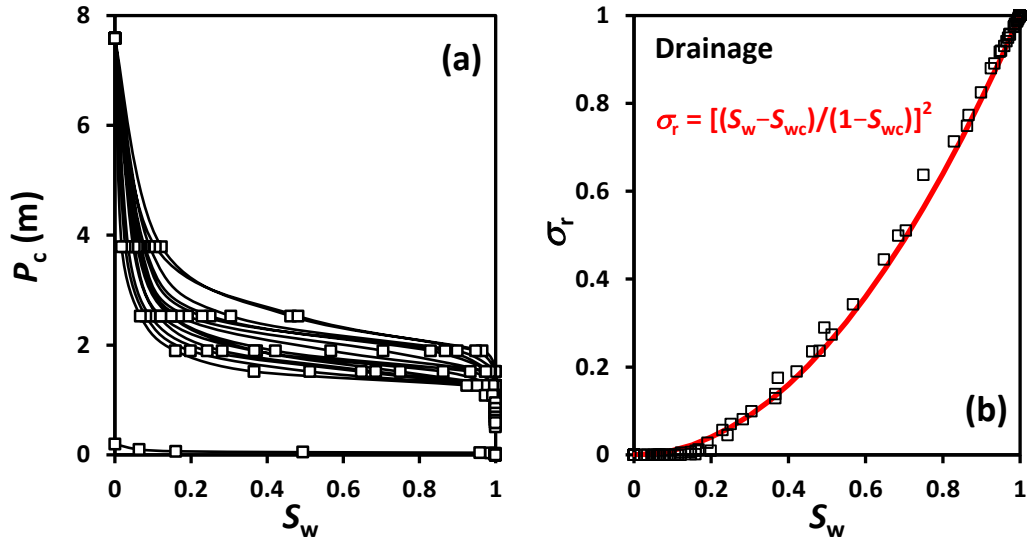

Figure 1. Dependence of (a) the capillary pressure $P_c$, and (b) the relative electrical conductivity $\sigma_r$ on the water saturation $S_w$ during drainage for the 15 packings of Mawer et al. (2015). The red line represents the predictions of Eq. (1) with a zero saturation threshold ($S_{wc} = 0$), negligible surface conductivity ($\sigma_s = 0$) and the conductivity exponent $t = 2$, while open squares denote the simulation results of Mawer et al. (2015).

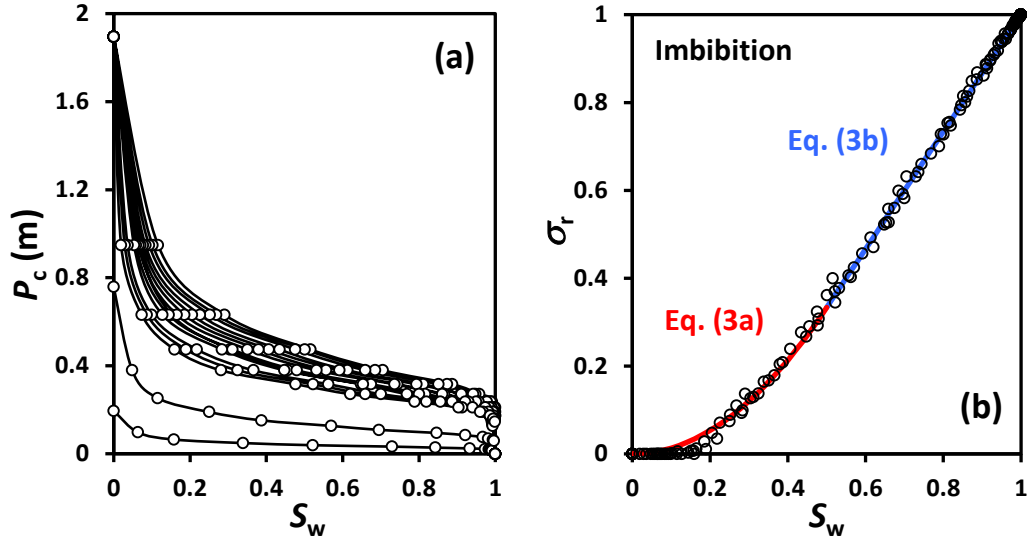

Figure 2. Dependence of (a) the capillary pressure $P_c$, and (b) the relative electrical conductivity $\sigma_r$ on the water saturation $S_w$ during imbibition in the 15 packings of Mawer et al. (2015). The red line represents Eq. 3(a), the universal power law with a zero saturation threshold ($S_{wc} = 0$), while the blue line denotes the EMA predictions, Eq. 3(b), with an average connectivity of $Z = 8$. Open circles are simulation data of Mawer et al. (2015). The crossover point above which Eq. 3(a) switches to Eq. 3(b) occurs at $S_{wx} = 0.5$.

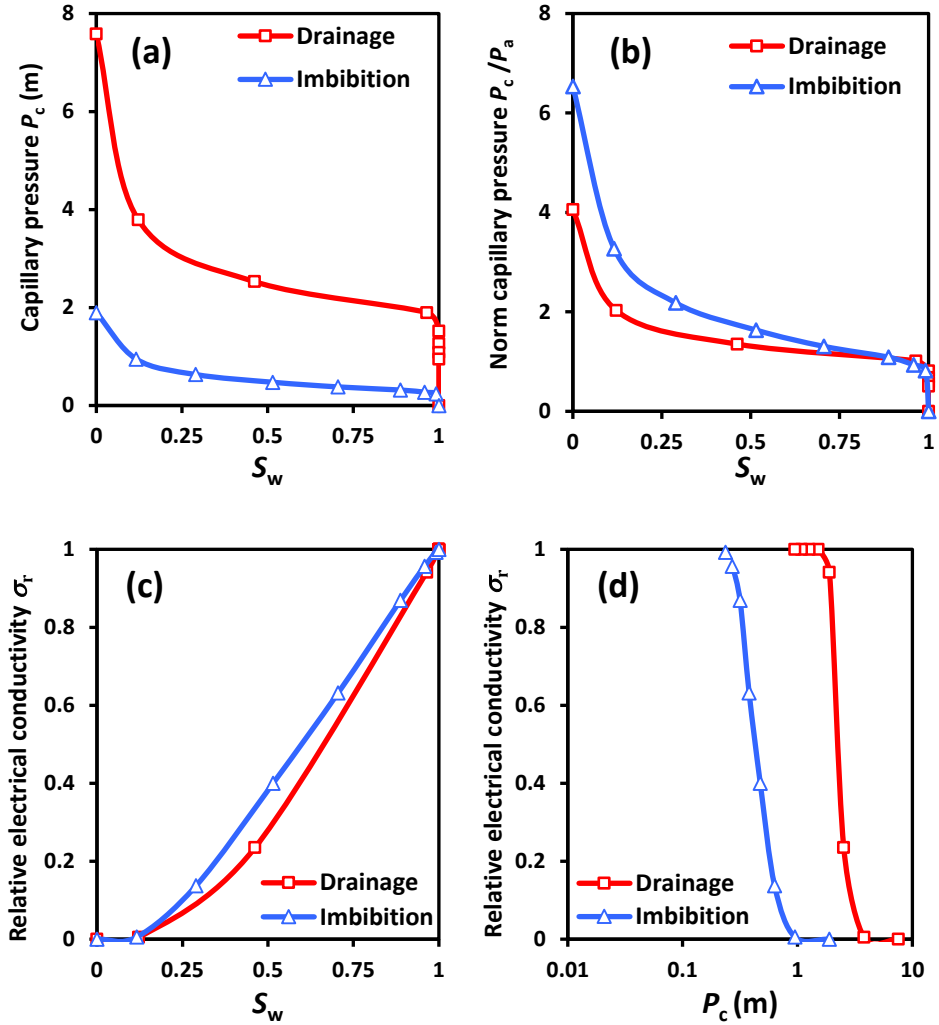

Figure 3. Dependence of (a) the capillary pressure $P_c$; (b) the normalized capillary pressure $P_c/P_e$, and (c) the relative electrical conductivity $\sigma_r$ on the water saturation $S_w$. (d) Dependence of the relative electrical conductivity on the capillary pressure during draiange and imbition for packing 3. Open triangles with the blue line represent the imbition process, while open squares with the red line show the drainage data of Mawer et al. (2015).

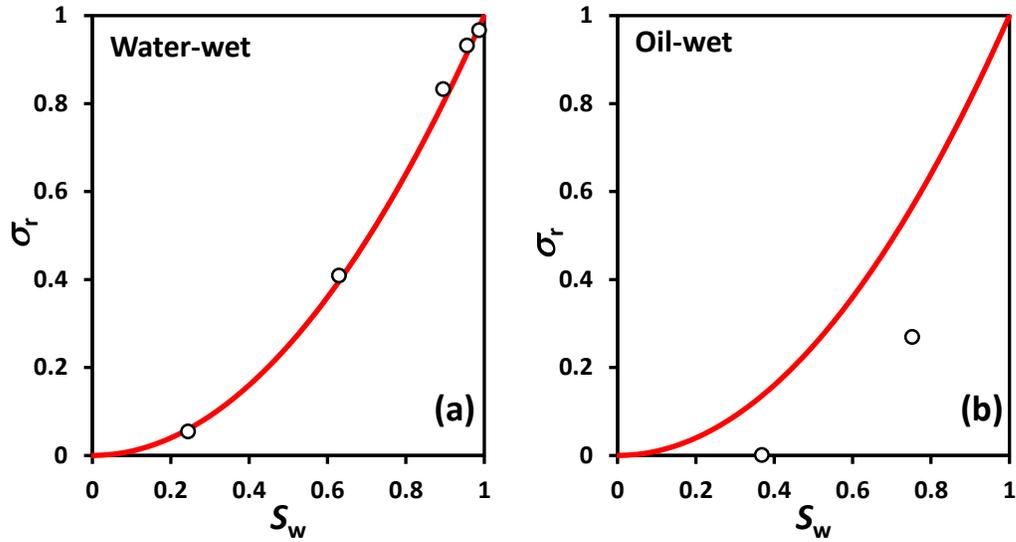

Figure 4. Dependence of the relative electrical conductivity $\sigma_r$ on the water saturation $S_w$ during the draiange process in the packings of glass beads under, (a) water- , and (b) oil-wet conditions reported by Knackstedt et al. (2007). Open circles denote the simulation results computed by using the image of the glass beads, while the red line represents the predictions of Eq. (1) with a zero saturation threshold ($S_{wc} = 0$), negligible surface conductivity ($\sigma_s = 0$), and conductivity exponent $t = 2$.

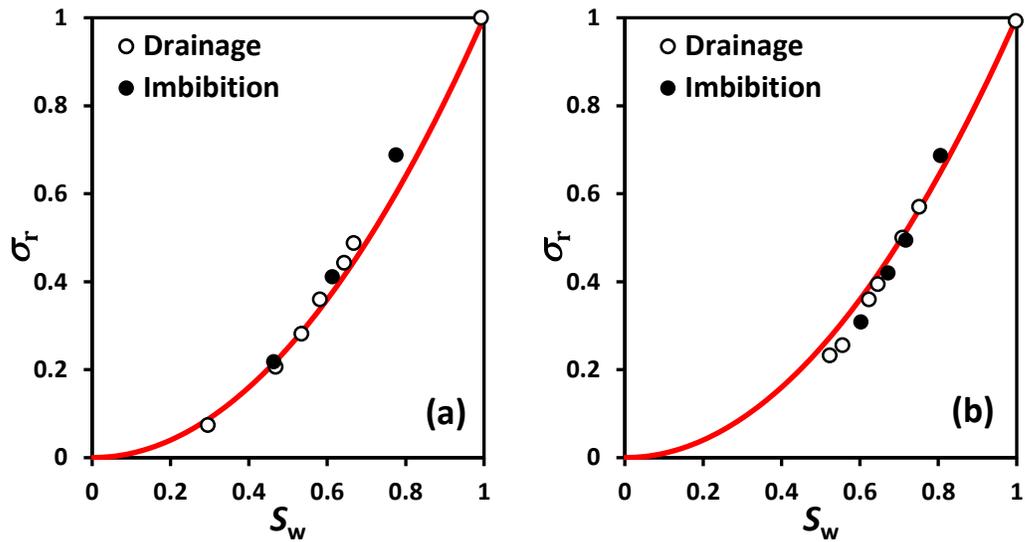

Figure 5. Dependence of the relative electrical conductivity $\sigma_r$ on the water saturation $S_w$ with, (a) n-decane, and (b) naphtha as the non-wetting phase in a water-wet packing of uniform glass beads reported by Sharma et al. (1991). Open and solid circles denote, respectively, the experimental measurements during drainage and imbibition. Red line represents the predictions of Eq. (1) with a zero saturation threshold ($S_{wc} = 0$), negligible surface conductivity ($\sigma_s = 0$), and the conductivity exponent $t = 2$.